# Gallium Arsenide Thermal Conductivity and Optical Phonon Relaxation Times from First-Principles Calculations

Tengfei Luo[1(a)], Jivtesh Garg[2], Junichiro Shiomi[3], Keivan Esfarjani[2] and Gang Chen[2(b)]

[1] Aerospace and Mechanical Engineering, University of Notre Dame, Notre Dame, 46556, USA
[2] Mechanical Engineering, Massachusetts Institute of Technology, Cambridge, 02139, USA
[3] Mechanical Engineering, The University of Tokyo, Tokyo, 113-8656, JAPAN

**Abstract** – In this paper, thermal conductivity of crystalline GaAs is calculated using first-principles lattice dynamics. The harmonic and cubic force constants are obtained by fitting them to the force-displacement data from density functional theory calculations. Phonon dispersion is calculated from dynamical matrix constructed using the harmonic force constants and phonon relaxation times are calculated using Fermi's Golden rule. The calculated GaAs thermal conductivity agrees well with experimental data. Thermal conductivity accumulations as a function of phonon mean free path and as a function of wavelength are obtained. Our results predict significant size effect on the GaAs thermal conductivity in the nanoscale. Relaxation times of optical phonons and their contributions from different scattering channels are also studied. Such information will help understanding hot phonon effects in GaAs-based devices.

**Introduction**

Gallium arsenide (GaAs) is an important semiconducting material that has been widely used in optoelectronic and microelectronic devices. Excessive temperature rise often is the limiting factor for the device performance and reliability. Although GaAs thermal conductivity values have been measured,[1] detailed thermal transport in GaAs has not been well studied especially at the nanoscale where finite size effect is expected. To accurately predict the size effect on the GaAs thermal conductivity that is presented at the nanoscale, mean free path (MFP) of individual phonon modes and their contributions to the thermal transport need to be evaluated. In many GaAs-based optoelectronic devices, hot phonon effects, which refer to the phenomena that optical phonons can have highly non-equilibrium population when absorbing energy from photoexcited electrons,[2,3] are important for their performance. The hot phonon effect is a strong function of the intrinsic relaxation times of the optical phonons as their energy needs to be relaxed to acoustic phonons due to three-phonon interactions. As a result, determining these relaxation times of optical phonons is critical to understanding and estimating the hot phonons effects in GaAs based devices.

In this work, we use first-principles density functional theory calculations to extract the harmonic and cubic force constants using a fitting procedure established by Esfarjani and Stokes.[4] The force constants are then used for thermal conductivity evaluations using lattice dynamics. This method and similar methods have been successfully used to calculate the thermal conductivity of a number of semiconducting materials.[5-10] Phonon group velocity and heat capacity are obtained from lattice dynamics using the harmonic force constants. Using the cubic force constants, the relaxation times of different phonon modes are evaluated using Fermi's Golden rule under the single mode relaxation time approximation. Thermal conductivity is calculated and the contributions of phonons of different MFPs and wavelengths to the total thermal conductivity are obtained. We also examined the relaxation times of optical phonons at high symmetry points and their decompositions into different scattering channels.

**Method**

The lattice thermal conductivity ($\kappa$) of a crystal can be calculated using the phonon Boltzmann transport equation which yields an expression shown in Eq. (1).

$$\kappa = \frac{1}{3VN_k}\sum_{k\lambda} c_{k\lambda} \mathrm{v}_{k\lambda}^2 \tau_{k\lambda} = \frac{1}{3VN_k}\sum_{k\lambda} c_{k\lambda} \mathrm{v}_{k\lambda} \Lambda_{k\lambda} \quad (1)$$

The subscript $k$ refers to the wave-vector in the first Brillouin zone and $\lambda$ refers to different phonon branches. $V$ is the volume of the unit cell, $N_k$ is the number of discrete $k$-points, $c$ denotes the heat capacity per mode derived from the Bose-Einstein distribution, $\mathrm{v}$ the group velocity, $\tau$ the phonon relaxation time, and $\Lambda$ the phonon mean free path ($\Lambda = \mathrm{v}\cdot\tau$). The first-principles density functional theory calculation is performed using the planewave based Quantum-Espresso package.[11] Ultrasoft pseudopotentials with Perdew-Zunger local density approximation[12] are used for both gallium and arsenic atoms. A planewave cut-off of 60 Rydberg and a Monkhorst-Pack[13] mesh of 4×4×4 in the *k*-space are chosen based on the convergence test of the lattice energy. In the fitting process to obtain the force constants, the cut-off of the harmonic force constants is the fifth nearest neighboring shell and that of the cubic force constants is set to the nearest neighboring shell. Such a choice is based on the convergence test of the phonon dispersion with respect to cut-off ranges. These cut-offs are used in all the thermal conductivity and phonon property calculations to minimize the computational time. Phonon dispersion relation is calculated using the harmonic force constant. The group velocity is obtained by taking the gradient of the phonon dispersion relation ($\mathrm{v}_{k\lambda} = \nabla_k \omega_{k\lambda}$). The phonon dispersion is calculated by the dynamical matrices which are constructed using the harmonic force constants. Non-analytical terms due to the Columbic forces are added to the dynamical matrices[14] with the Born charges ($Z_{Ga}$=2.105, $Z_{As}$=-2.105) and the dielectric constant ($\varepsilon$ =13.217) calculated from the density functional perturbation theory.[15] The masses used in the calculations are 69.723 and 74.922 for Ga and As, respectively. The phonon relaxation times are calculated using Fermi's Golden rule utilizing the cubic force constants.[16] Scatterings due to anharmonicity with orders higher than three are ignored since higher order anharmonic scattering are usually very weak at room and moderate temperatures.[17] More details of the Fermi's Golden rule calculation can be found in Ref. 6.

[a] E-mail: tluo@nd.edu
[b] E-mail: gchen2@mit.edu

**Results and Discussion**

The calculated GaAs phonon dispersion relation shown in Fig. 1a agrees well with experimental data,[18] which is an indication of the high accuracy of the harmonic force constants obtained from the first-principles calculations. It can be seen from the dispersion relation that the longitudinal acoustic (LA) modes have the highest group velocities near the Brillouin zone center indicated by the largest slopes around the Γ point. The two transverse acoustic (TA) branches also have high group velocities, but those of the optical modes are relatively small as reflected by the flatness the optical dispersion curves.

To verify the accuracy of the cubic force constants, the mode Grüneisen parameters, which are indications of the anharmonicity of a crystal, are calculated and compared to experimental data. The calculated mode Grüneisen parameters agree well with the available reference data (Fig. 1b).[19-21] The small discrepancies between the calculation and experiment should be the consequences of the finite cutoff of the neighboring shells in the fitting of the force constants.[6] We have calculate the Gruneisen parameters when including the 2$^{nd}$ and 3$^{rd}$ nearest neighbors for the cubic constant fitting. However, when more neighbors are included, the Grüneisen parameters of the two TA modes will be smaller in magnitude than the experimental data but this effect is only obvious near the zone boundaries (X and L points). The Grüneisen parameters near the zone centers do not show significant neighbor shell dependence. This should not result in large errors in thermal conductivity calculations since the phonon group velocities near these regions are very small. The LA mode also shows shifts when more neighbor shells are included, but we found that using the nearest neighbor shell for cubic constant fitting results in values closer to experiments. The optical modes are not much influenced when more neighbors are included.

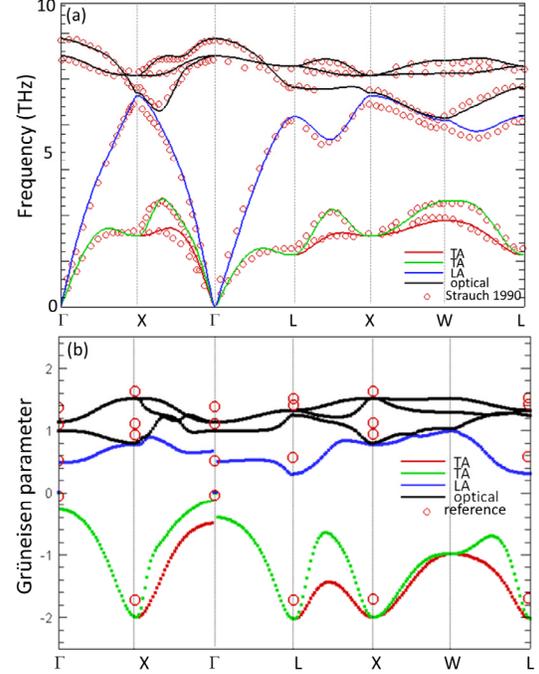

Figure 1 (a) Phonon dispersion of GaAs calculated using first-principles and from Ref.18; (b) Mode Grüneisen parameters from first-principles and from various references.[19-21]

The two TA branches have the smallest Grüneisen parameters near the Brillouin zone center indicating small anharmonicities of these modes. Near the Brillouin zone edges, TA modes become very anharmonic as indicated by the large Grüneisen parameters near X and L points. Over the whole Brillouin zone, LA modes have relatively small Grüneisen parameters and those of the optical modes are larger. Larger Grüneisen parameters indicate higher anharmonicities which generally lead to greater scatterings and thus shorter relaxation times. Considering the aforementioned observations on the group velocities and anharmonicities, the contribution from the acoustic modes to the thermal conductivity is expected to be much larger than that from the optical modes.

With the confidence on the force constants, the mode relaxation times are calculated using Fermi's Golden rule with both normal and umklapp scatterings included. Figure 2a and 2b show the relaxation times due to normal and Umklapp scatterings at 300K, respectively. It is seen that for low frequency modes normal scattering relaxation times scale with $\omega^{-2}$ and Umklapp scattering relaxation times scale with $\omega^{-3}$, which is consistent with that found in silicon.[6] The dashed lines shown in Fig. 2a and 2b which represent these frequency dependencies generally coincide well with the discrete data

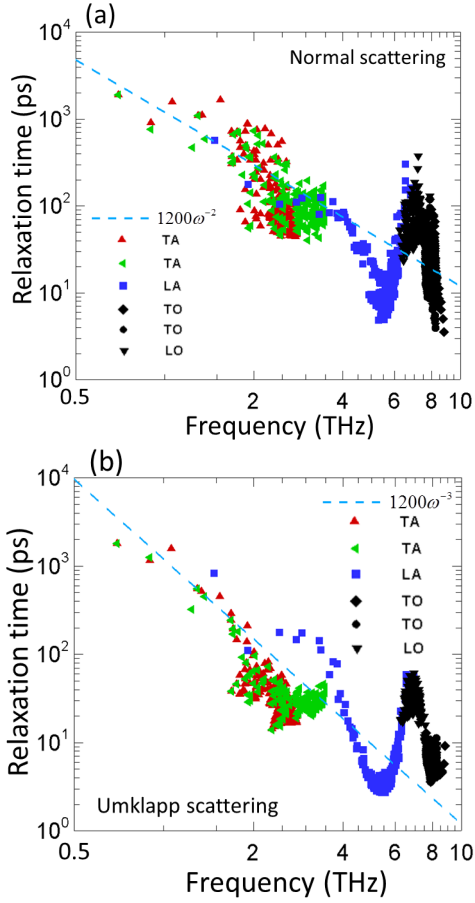
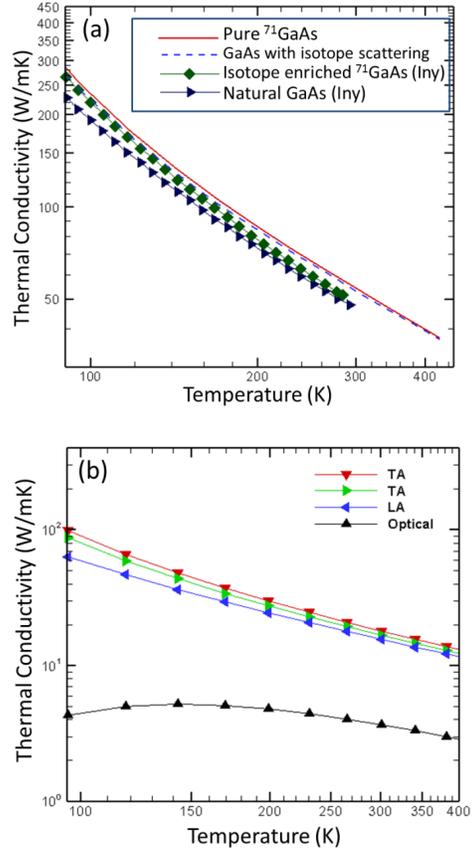

Figure 2 phonon relaxation times at 300K of different branches as a function of frequency due to (a) normal scattering and (b) Umklapp scattering. The dashed line in panel (a) represents a $\tau = 1200\omega^{-2}$ relation and that in panel (b) shows a $\tau = 1200\omega^{-3}$ relation.

from the lattice dynamics calculations in the low frequency region. In general, it is observed that the acoustic modes have much longer relaxation times than those of the optical modes. Such an observation is consistent with trends expected from the calculated mode Grüneisen parameters. The relaxation times of LA modes span a large range, and their relaxation times become comparable to those of optical modes at high frequency region (>4THz). At the low frequency region, the relaxation times of LA are comparable to those of the two transverse acoustic branches.

Combining the modal heat capacities, group velocities and relaxation times obtained from the first-principles calculations, thermal conductivity of GaAs can be calculated using Eq. (1). The first-principles results are from phonon calculations using a 14×14×14 grid in the k-space. We have confirmed that the thermal conductivity is converged with respect to grid size when the 14×14×14 grid is used. To make our calculation results comparable to experiments, we used 71 as the mass of the gallium atom. We found that the thermal

Figure 3 (a) $^{71}$GaAs thermal conductivity and that with isotope scattering from first-principles calculations and from Ref. 1; (b) Thermal conductivity contribution from acoustic branches and optical branches.

conductivities of $^{71}$GaAs and $^{69.723}$GaAs are almost the same (with a difference of less than 0.4%). Compared to the measured thermal conductivity of isotope enriched $^{71}$GaAs, the first-principles values are uniformly larger (Fig. 3a). The discrepancy is about 7% at low temperatures and increases to ~10% at around 300K. However, considering that there might be defects in the measurement sample, it is understandable that its thermal conductivity is lower than our prediction for a perfect GaAs crystal. Another possible factor that can contribute to such discrepancy is the higher order phonon scattering that is not included in our calculation. Such scattering is more pronounced at higher temperatures. However, we believed that a prediction within 10% of the experimental value is very good considering the above-mentioned uncertainties.

The measured thermal conductivity of natural GaAs, which contains isotopes, is further lower than the first-principles predictions by ~20% and ~14% at low and room temperatures, respectively. To study possible isotope scattering effects, we employed the Rayleigh scattering formula ($\tau_{iso}^{-1} = A\omega^4$, where $A = 2.95 \times 10^{-45} s^3$ calculated from the formula given in Ref. 22 using an isotope concentration given in Ref. 1) to include the contribution of isotope scattering to the thermal conductivity. Such isotope scattering results in a decrease in thermal conductivity by about 2% at room temperature. This, however, does not completely offset the larger discrepancy (14% at room temperature) between our calculation results and that from experiments on the natural GaAs. It is possible that other defects such as dislocations also contribute to phonon scattering.

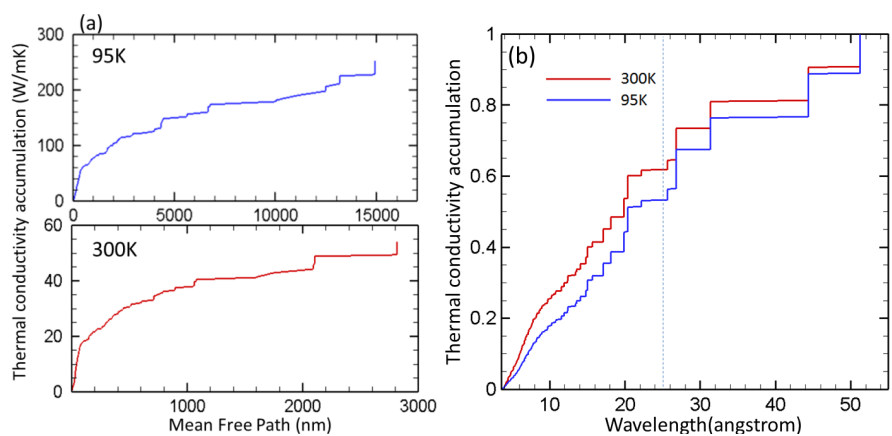

Figure 4 Accumulative thermal conductivity with respect to (a) phonon mean free path and (b) phonon wavelength at 95K and 300K.

We further decompose the thermal conductivity into the contribution from the three acoustic branches and the optical branches (Fig. 3b). It is seen that the contribution from the optical branches is about one order of magnitude smaller than that from any of the three acoustic branches. The three acoustic branches contribute almost the same amount of thermal conductivity. At 300K, the thermal conductivity contributed from the two TA modes are 18.1, 16.9 W/mK, respectively, and that from the LA mode is 15.7 W/mK. The three acoustic modes together contribute over 90% of the total thermal conductivity over the whole temperature range.

The first-principles based method can also provide more detailed information beyond the thermal conductivity values. According to Eq. (1), the total thermal conductivity can be decomposed into the contributions from individual phonon modes which can be sorted according to their MFP.[23, 24] This information is important since it indicates how phonons with different MFPs contribute to the total thermal conductivity. This offers valuable guidelines for predicting the size effect in small scale materials. Figure 4a shows the thermal conductivity accumulation with respect to phonon MFP at 300K calculated on a 14×14×14 *k*-point grid. It is seen that the phonon MFP in GaAs spans more than three orders of magnitude and about half of the total thermal conductivity is contributed from phonons with MFPs longer than 350nm. Such wide mean free path distribution suggests strong size effects on phonon transport in optoelectronic devices [25,26] Since phonons will have longer MFP at lower temperatures due to reduced phonon scatterings, the size effect on thermal conductivity will be more significant. Figure 4a also shows the accumulative thermal conductivity at 95K. At this temperature, the phonon MFP in GaAs spans more than four orders of magnitude and half of the thermal conductivity is contributed from phonons with MFP longer than 4 $\mu m$. Such information, which predicts the finite size effects on nanostructured GaAs thermal conductivity, is certainly very useful for thermal management design in GaAs based micro-/nano-electronic devices.

It is also of interest to investigate how phonons with different wavelength contribute to the thermal conductivity since they have different transmissivity across interfaces, with long wavelength phonons usually having higher and specular transmissivities. Figure 4b shows the thermal conductivity accumulation with respect to the phonon wavelength. The longest wavelength available in the 14x14x14 grid calculation is about 50 Å. If we take the half of the longest wavelength of 25 Å as a dividing point, we can see that phonons with wavelength longer than this point contribute ~40% and ~50% at 300K and 95K, respectively. Unlike the phonon MFP dependent thermal conductivity, the thermal conductivity accumulation profiles with respect to the phonon wavelength at low and high temperatures are not significantly different.

We calculated the inverse relaxation times of optical phonons at high symmetry points and decompose the overall scatterings into the contributions from different scattering channels. Figure 5 shows the inverse relaxation times and their decompositions of optical phonons at different symmetry points. The values of the LO mode at the Γ points (Fig. 5a) are uniformly lower than previous first-principle calculations[27] but agree reasonably well with experimental values.[28, 29] The inverse relaxation times of the TO mode at the Γ points are closer to previous calculation results.[27, 30] It is seen that the major energy decay channel of both the LO and TO modes at all points studied is via scattering to two acoustic modes (op-ac-ac). The only exception is one of the TO modes at the W point (Fig. 5c – middle panel). For this mode, the contribution of the op-ac-op processes, in which the optical phonons is scattered by one acoustic mode and one optical mode, is comparable to that of the op-ac-ac processes.

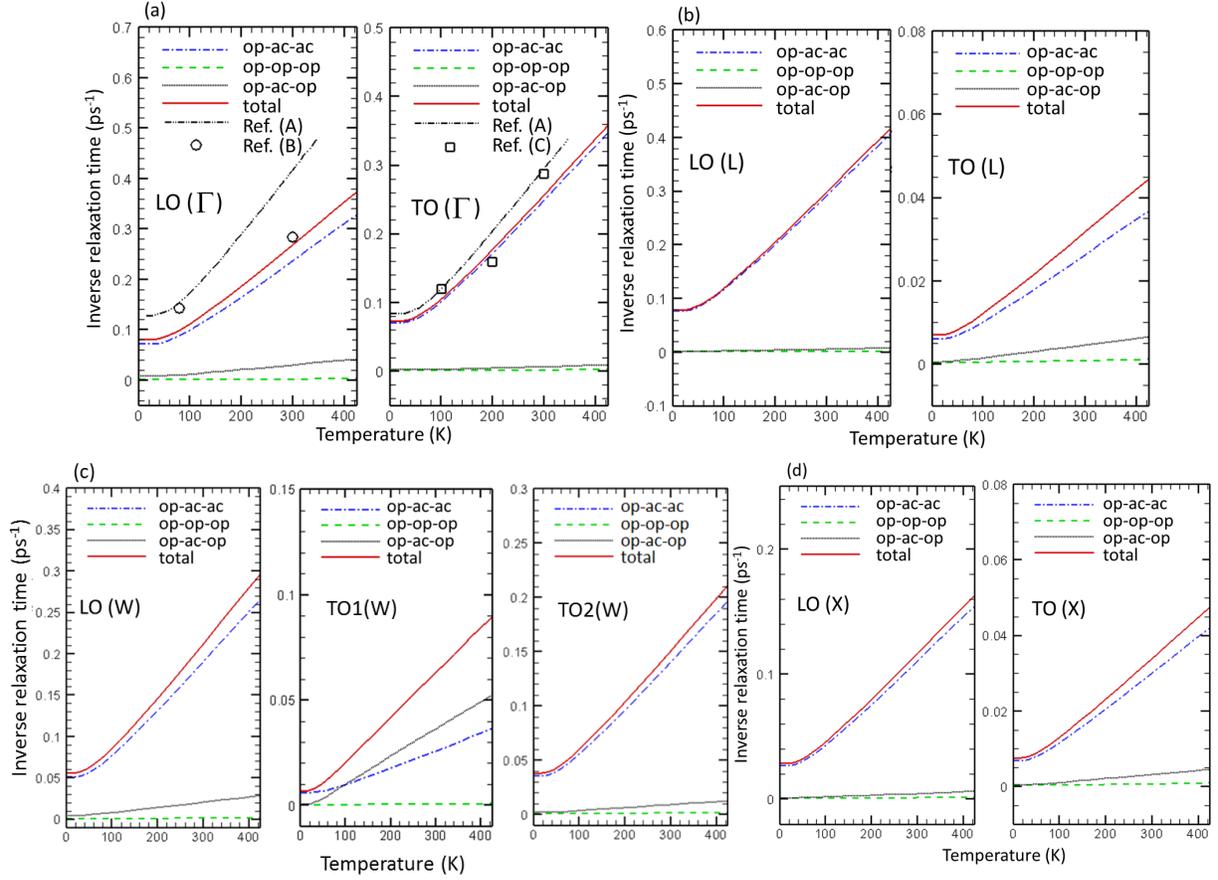

Figure 5 Inverse relaxation times of optical phonons and their decompositions at (a) $\Gamma$ point, (b) L point, (c) W point and (d) X point. (in panel (a): Ref. (A): first-principles calculation from Ref.27; Ref. (B): experimental results from Refs. 28, 29; Ref. (C): first-principles calculation from Ref.30)

Scattering involving only optical modes (op-op-op) is prohibited at all points due to the energy and momentum selection rules required for three phonon scattering, and thus contribution from such scattering channels is negligible. Such an observation is identical to previous calculations.[27] The op-ac-op processes generally contribute less than 10% of the total scatterings at 300K with an exception seen in one of the TO modes at the W point. Majority of the modes studied have inverse relaxation times ranging from 0.1 to 0.3 ps$^{-1}$ at 300K, corresponding to relaxation times of ~3-10 ps. However, the TO modes at the L and X points have much longer relaxation times (~30 ps).

**Summary**

In summary, we used a first-principles lattice dynamics method, which does not need empirical inputs or fitting parameters, to calculate the thermal conductivity of GaAs. The calculated thermal conductivities are uniformly higher than the experimental data on isotope enriched $^{71}$GaAs and natural GaAs samples by 7-10% and 14-20%, respectively, over the temperature range of 95-400K. We regard such agreement to be good considering the defect scatterings in the measurement samples, in addition to our limitation to three phonon scattering. We also studied the thermal conductivity accumulation as a function of phonon MFP which indicates the contributions of phonons with different MFP to the overall thermal transport in bulk GaAs. Results presented demonstrate that significant size effects can present in nanoscale GaAs materials even at room temperature. Optical phonon relaxation times are investigated at several high symmetry points and the contribution from different scattering channels are studied. Detailed information provided in this paper offers valuable guidance on the thermal management and hot phonon analysis of GaAs based optoelectronic and microelectronic devices.

***

T.L. thanks Ms. Zhiting Tian for useful discussions. This research was supported in part by the Solid State Solar-

T. Luo *et al.*

Thermal Energy Conversion Center (S3TEC), an Energy Frontier Research Center funded by the U.S. Department of Energy, Office of Science, Office of Basic Energy Sciences under Award Number DE-SC0001299 (J.G., K.E., and G.C.), DARPA NTI program via Teledyne (T.L.), XSEDE resources provided by TACC Ranger and SDSC Trestles under grant number TG-CTS100078. T.F. thanks the startup fund from the University of Notre Dame.